\documentclass{iau}

\usepackage{amsmath}
\usepackage{graphicx}
\usepackage{multirow}

\begin{document}

\lefttitle{Cambridge Author}
\righttitle{Proceedings of the International Astronomical Union: \LaTeX\ Guidelines for~authors}

\jnlPage{1}{7}
\jnlDoiYr{2021}
\doival{10.1017/xxxxx}

\aopheadtitle{Proceedings IAU Symposium}
\editors{C. Sterken,  J. Hearnshaw \&  D. Valls-Gabaud, eds.}

\title{Dynamical evolution modeling of the Collinder 135 & UBC 7 binary star cluster}

\author{
Marina Ishchenko1
\and
Dana Kovaleva2
\and
Peter Berczik4,3,1
\and 
Nina Kharchenko1
\and
Anatoly Piskunov2
\and
Evgeny Polyachenko2
}

\affiliation{\institute{Main Astronomical Observatory, National Academy of Sciences of Ukraine, 27 Akademika Zabolotnoho St, 03143 Kyiv, Ukraine			
           \and
           Institute of Astronomy, Russian Academy of Sciences, 48 Pyatnitskya St, Moscow 119017, Russia   
           \and
           National Astronomical Observatories and Key Laboratory of Computational Astrophysics, 
		   Chinese Academy of Sciences, 20A Datun Rd., Chaoyang District, Beijing 100101, China
           \and
           Zentrum f\"ur Astronomie der Universit\"at Heidelberg, Astronomisches Rechen-Institut, M\"onchhofstr 12-14, 69120 Heidelberg, Germany
           \and
           Landessternwarte, Zentrum f\"ur Astronomie der Universit\"at Heidelberg, K\"onigstuhl 12, 69117, Heidelberg, Germany}           }
           
\begin{abstract}
The purpose of the present work is a detailed investigation of the dynamical evolution of Collinder 135 and UBC 7 star clusters. We present a set of dynamical numerical simulations using realistic star cluster N-body modelling technique with the forward integration of the star-by-star cluster models to the present day, based on best-available 3D coordinates and velocities obtained from the latest Gaia EDR3 data release. 
\end{abstract}


\maketitle

\section{Introduction}

Recently, based on analysis of data of Gaia space mission \cite{2016A&A...595A...1G}, several pairs of star cluster joint by common origin were discussed \citet{2021MNRAS.503.5929B,  2020ApJ...900L...4P, 2019A&A...624A..34Z, 2020A&A...642L...4K}. Collinder~135 (hereafter Cr~135) and newly discovered \cite{2018A&A...618A..59C} UBC~7 are located in the Vela-Puppis region which attracts active attention since Gaia data allowed to appreciate its complicated history of evolution, space and kinematic structure \citet{2018MNRAS.481L..11B, 2019A&A...621A.115C, 2019A&A...626A..17C, 2020MNRAS.491.2205B}. We have  demonstrated \cite{2020A&A...642L...4K} that these two clusters space location and velocities suggest that they might have been closer to each other at their initial history, 40 to 50 Myr ago. Backward orbital integration indicates also that the clusters might have been gravitationally bound in the past if they were significantly more massive before their violent relaxation.  

We used Gaia DR2 and EDR3 data to restore the probable members of Cr~135 and UBC~7 in the full possible scope, adjusting method by \cite{2012A&A...543A.156K}. Based on obtained dataset, we constrain present day parameters of the two clusters, such as space positions, space velocities, masses and density profiles. These data are used further as boundary conditions for dot-to-dot numerical simulation of dynamical evolution of this pair of star clusters.

\section{N-body modelling of the star clusters evolution}
\label{sec:dyn}

\subsection{Numerical method}

We used the \PGPU code for the numerical solution of the equations of motion. 
The \PGPU package uses a high order Hermite integration scheme and individual block time steps (the code supports time integration of particle orbits with schemes of 4$^{\rm th}$, 6$^{\rm th}$ and even 8$^{\rm th}$ order). Such a direct $N$-body code evaluates in principle all pairwise forces between the gravitating particles, and its computational complexity scales asymptotically with $N^2$. We refer the more interested readers to a general discussion about different $N$-body codes and their implementation in \cite{spurzem2011a, spurzem2011b}. 

The \PGPU code is fully parallelized using the MPI library. This code is written from scratch in {\bf \tt C++} and is based on earlier CPU serial $N$-body code (YEBISU; \cite{NM2008}). The MPI parallelization was done in the same {\em j} particle parallelization mode as in the earlier \PGRAPE code \cite{HGM2007}. The current version of the \PGPU\footnote{\tt ftp://ftp.mao.kiev.ua/pub/berczik/phi-GPU/} code uses a native GPU support and direct code access to the GPU using the NVIDIA native CUDA library. The multi GPU support is achieved through global MPI parallelization. Simultaneously, our code effectively exploits also the current CPU's OpenMP parallelization. The code is designed to use different softening parameters for the gravity calculation (if it is required) for different astrophysical components in our simulations like SMBHs, dark matters or stars particles. More details about the \PGPU code public version and its performance we are presented in \cite{SBZ2012, BSW2013}.
The present code is well tested and has already been used to obtain important results in our earlier large scale (up to few million body) simulations \cite {2016MNRAS.460..240K}, \cite{2012ApJ...748...65L} and \cite{2014ApJ...792..137Z}.

\subsection{Orbits of Cr 135 and UBC 7}

We integrated the center of mass orbits both star clusters forwards in time in the analytic Milky Way potential. The integration time (50 Myr) was equal to the most probable age of the Cr 135 and UBC 7 (link). The full 3D orbits of the evolution are shown in Fig.~\ref{fig:orb}. The orbits integration with a simple integrator yielded the initial position of the Cr 135 ans UBS 7 at the time of their formation (see Table~\ref{tab:data1}). From the initial positions of the clusters we ran full N-body models up to the present time. It should be noted that the clusters rotate around each other. The present-day position and velocity of the Cr 135 and UBC 7 obtained from numerical simulation are given in Table~\ref{tab:data2}.

\begin{table*}[tbp]
\setlength{\tabcolsep}{4pt}
\centering
\caption{Initial position and velocity values for Cr 135 and UBC 7 center of mass in Cartesian Galactic coordinates. Taken from {\tt \#(53,61)} model, Paper I.}
\label{tab:data1}
\begin{tabular}{ccccccccccc}
\hline 
\hline 
Cluster & $X$, pc & $Y$, pc & $Z$, pc & $V_x$, km/s & $V_y$, km/s & $V_z$, km/s \\
\hline
Cr~135 & -1061.421 & -8382.545 & -22.70332 & -230.559 & 33.1769 & 5.52892\\
UBC~7 & -1065.137 & -8386.942 & -14.5731 & -230.792 & 34.0251 & 5.86780\\
\hline 
\end{tabular}
\vspace{6pt}
\end{table*}

\begin{table*}[tbp]
\setlength{\tabcolsep}{4pt}
\centering
\caption{Position and velocity values for Cr 135 and UBC 7 center mass in Cartesian Galactic coordinates at 50 Myr according to numerical simulation.}
\label{tab:data2}
\begin{tabular}{ccccccccccc}
\hline 
\hline 
Cluster & $X$, pc & $Y$, pc & $Z$, pc & $V_x$, km/s & $V_y$, km/s & $V_z$, km/s \\
\hline
Cr~135 & -8281.60 & -281.5499 & -3.59237 & -7.55748 & 237.672 & -4.95032\\
UBC~7 & -8283.713 & -251.8398 & -4.35904 & -7.23660 & 237.7669 & -4.983461\\
\hline 
\end{tabular}
\vspace{6pt}
\end{table*}

\begin{figure}
    \centering
    \includegraphics[scale=0.70]{orbit.eps}
    \caption{3D orbits evolution of the Cr 135 (rhombus-line) and UBS 7 (cross-line). The integration time is 50 Myr.}
    \label{fig:orb}
\end{figure}

The Cr 135 and UBS 7 orbits around the Galactic center is oriented in a clockwise direction as seen from the Galactic north pole. The orbits has a considerable vertical oscillation in the z-direction between $\pm$80 pc. In the radial direction the orbit is mildly eccentric and oscillates between $\pm$8 kpc. 

\subsection{Initial parameter space}
We present a numerical simulations using realistic star cluster N-body modelling by integrating a star-by-star cluster models to the present day. The code taking in to account the up to date stellar evolution models \cite{Banerjee2020}.  For the initial position and orbital velocities of the star clusters centre we used the selected  {\tt \#(53,61)} model from the our Paper I. The initial conditions for the clusters center of mass coordinates and velocity taken from the above selected (see  Table~\ref{tab:data1}). 

\begin{table}[htbp]
\setlength{\tabcolsep}{4pt}
\centering
\caption{Initial values of physical parameters for Cr 135 and UBC 7.}
\label{tab:king_w}
\begin{tabular}{cccccccc}
\hline 
\hline 
Cluster & $M_\odot$ & $N$ & $R$, pc & W$_{\rm 0}$ & \\
\hline
Cr~135 & 230 & 442 & 10 & 3 \\
UBC~7 & 200 & 384 & 7 & 11 \\
\hline 
\end{tabular}
\vspace{6pt}
\end{table}

We were looking for the best-fitting \cite{King_1966} 
models for the current observations from Gaia DR3. We assume that cluster age is 50 Myr. Our goal was to reproduce the final cumulative mass profile $M({\rm r}$). For the Cr 135 we used the range within $0 < r < 20$ pc and for UBC 7 -- $0 < r < 15$ pc. These limits corresponds to the clusters current Jacobi radius's. Because the initial masses of the clusters are quite uncertain, we used for the model fits the initial masses as a parameter. Since clusters are formed in molecular clouds with a low star formation efficiency, they are most probably supervirial after the gas expulsion phase  \cite{2021A&A...654A..53S}. For the initial mass function we used the \cite{2001MNRAS.322..231K} approximation, with the lower mass $m_{\rm l}$ = 0.1 $M_\odot$ and the upper mass $m_{\rm h}$ = 10 $M_\odot$ limits. The other two main parameters for the cluster initial models was a R$_{\rm core}$ and the King concentration parameter $W_{\rm 0}$, individually for the each clusters (see Table~\ref{tab:king_w}). For the stellar metallicity we used the value $Z = 2\%$ (assumed as a Solar value) for the both clusters. 

More than 50 individual models with stellar evolution have been computed. The total running time for one typical model on a AMD 3600X 4.1 GHz CPU with a GeForce RTX 2600 Super GPU card was about 2 min. Minimizing the difference between the cumulative number distribution of the observed clusters and the numerical model we find the best-fit parameters for both clusters, see Table~\ref{tab:king_w}.

\begin{figure*}[!htb]
    \centering
    \includegraphics[angle=-90,scale=0.30]{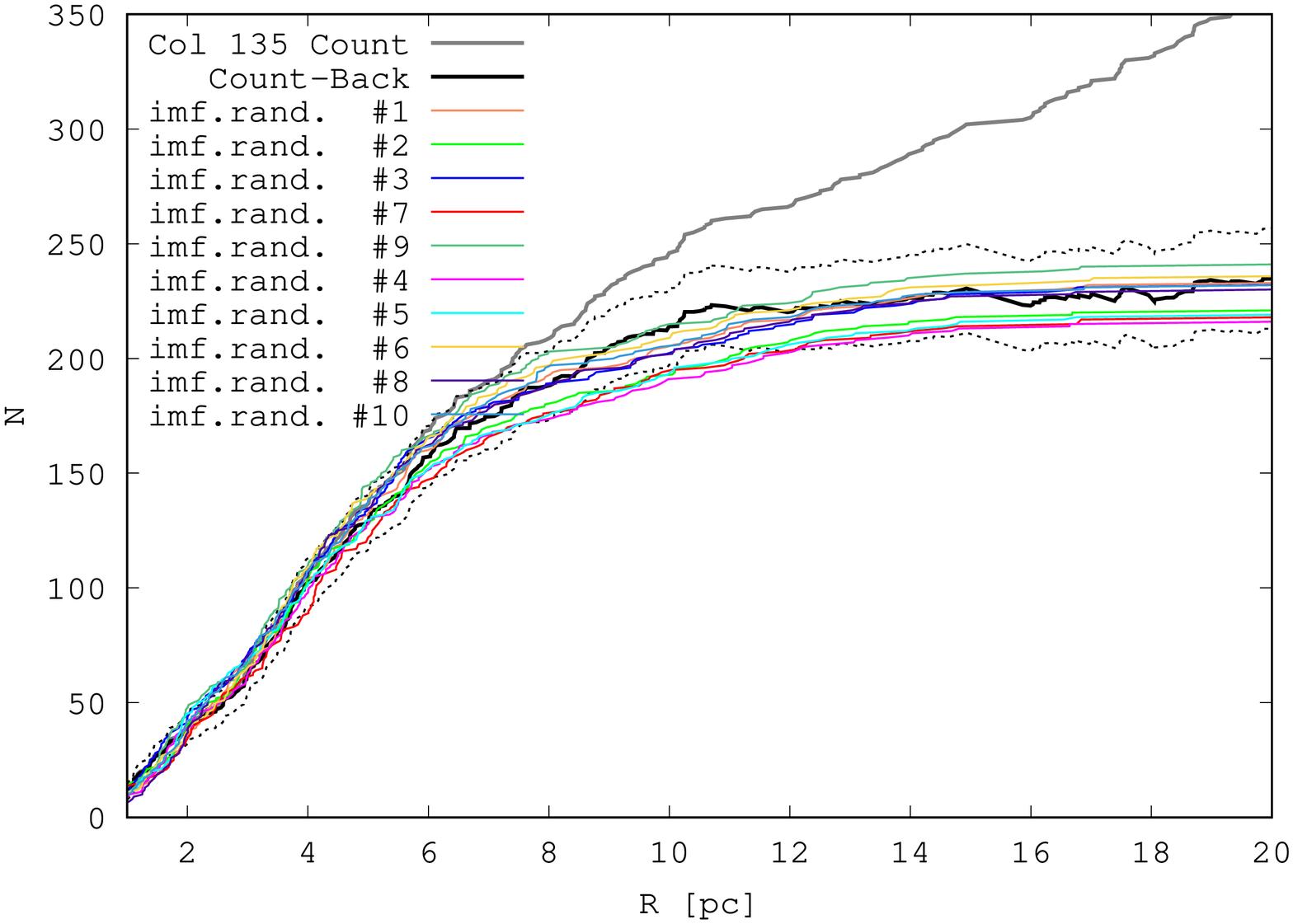}
    \includegraphics[angle=-90,scale=0.30]{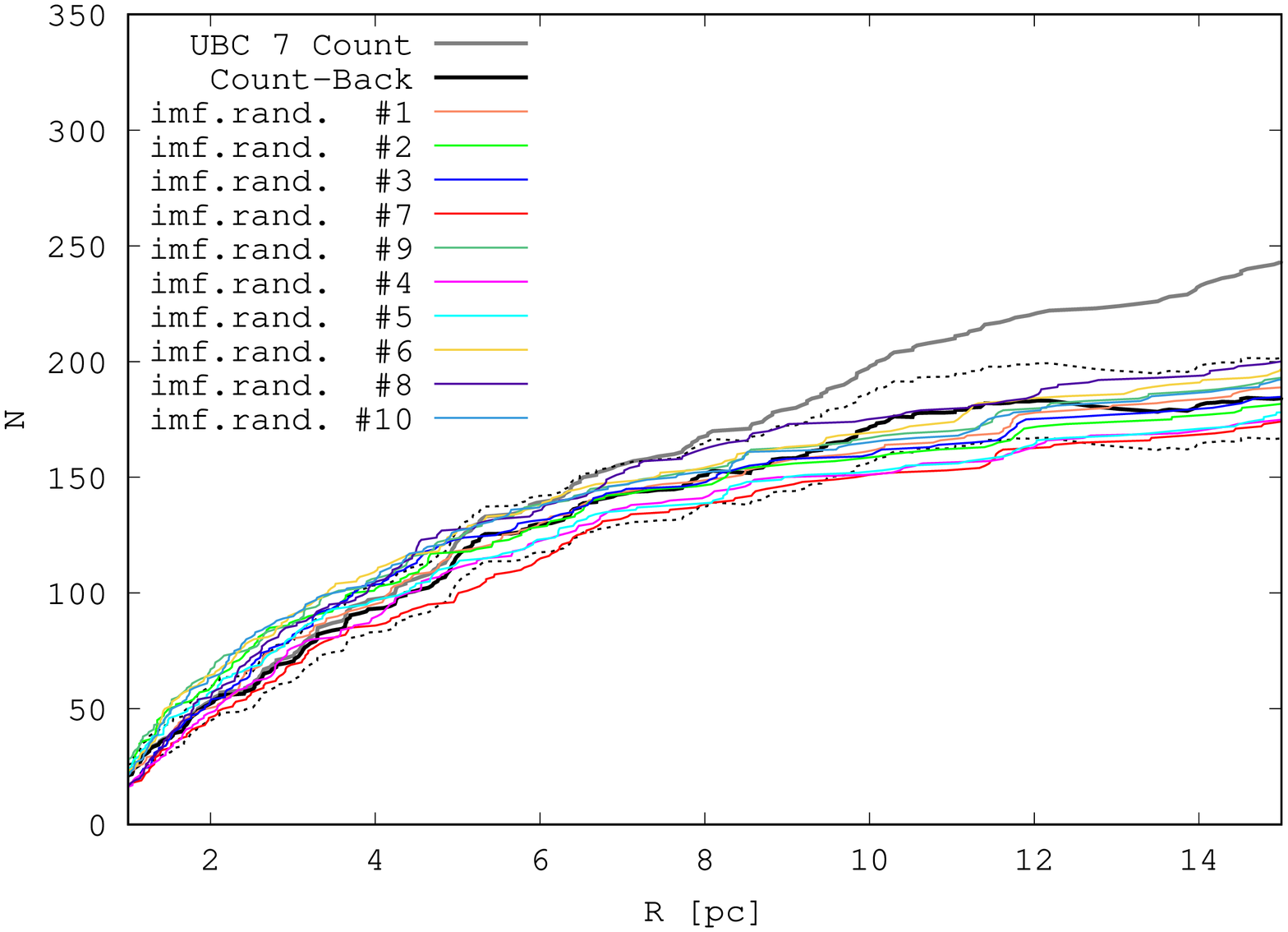}
    \caption{Cumulative number distribution of stars for clusters Cr 153 (left panel) and UBC 7 (right panel). The different color lines represents the set of randomization for initial mass function when the coordinates and velocities of the stars are fixed.} 
    \label{fig:imf-col}
\end{figure*}

\begin{figure*}[!htb]
    \centering
    \includegraphics[angle=-90,scale=0.30]{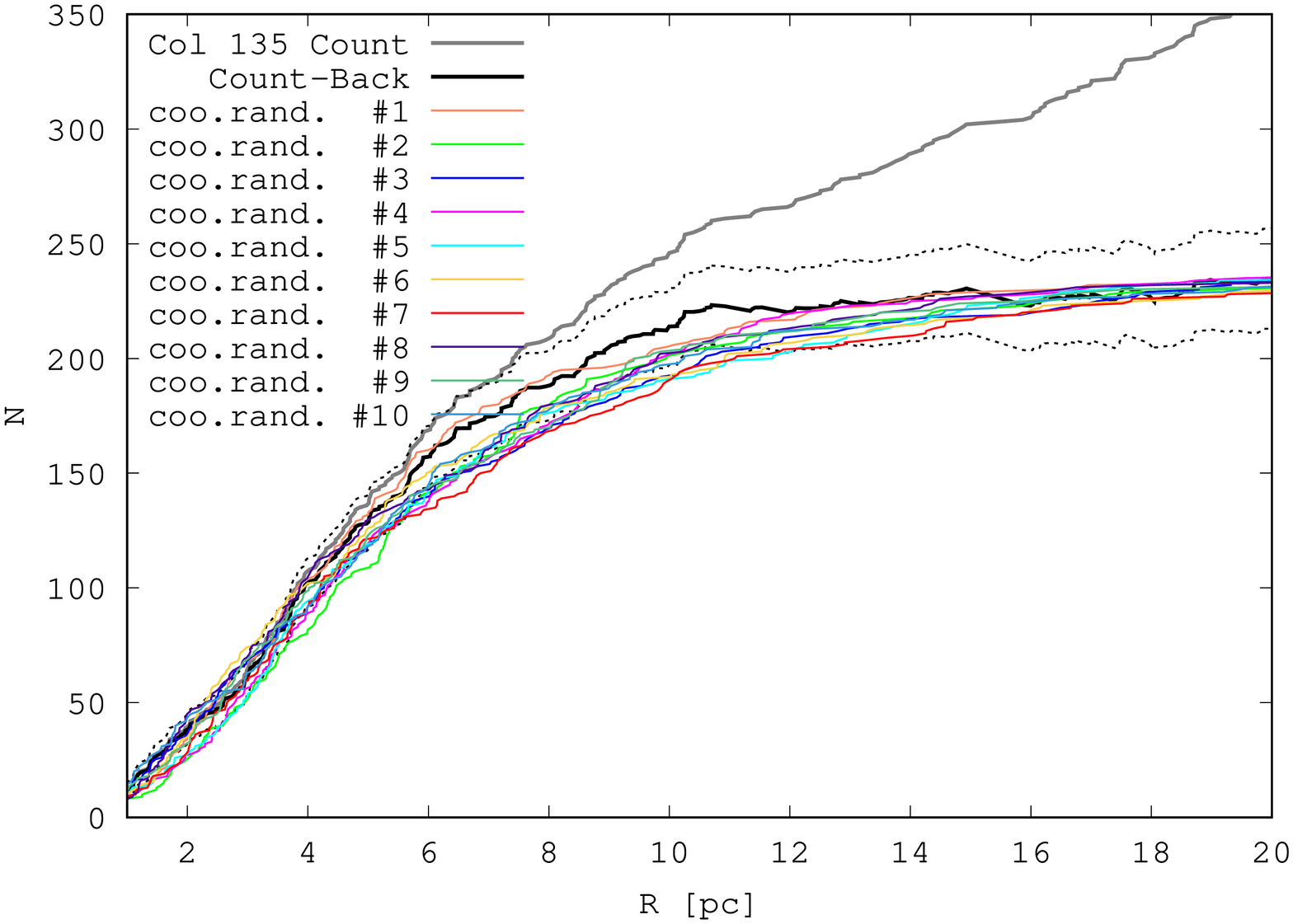}
    \includegraphics[angle=-90,scale=0.30]{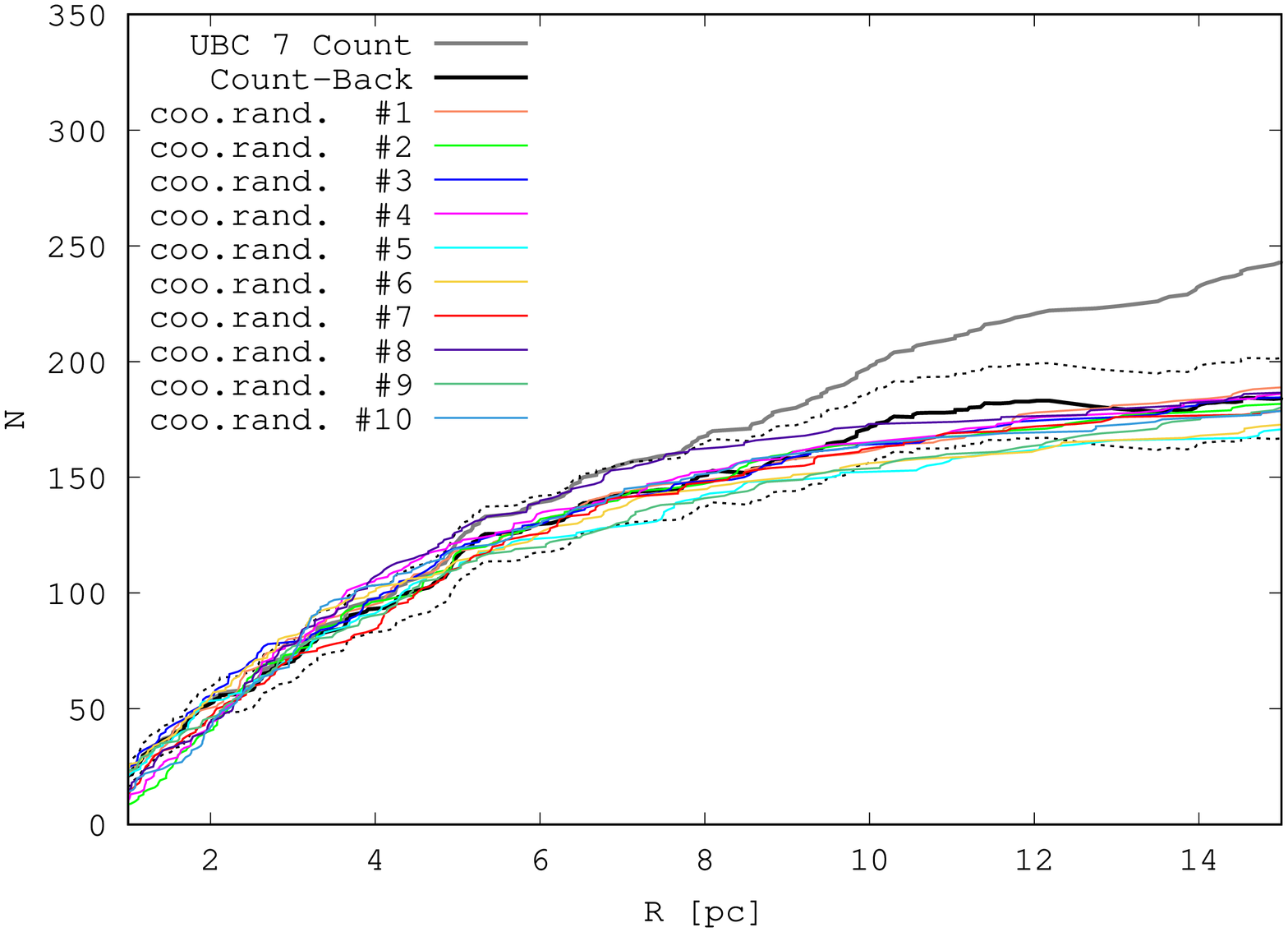}
    \caption{Same as Fig.~\ref{fig:imf-col} but randomization for initial coordinates and velocities was done with the fixed initial stellar mass function.}
    \label{fig:coo-col}
\end{figure*}

After this fitting procedure we run 20 more numerical models (with the clusters same physical parameters). First we generated 10 random sets with different initial mass function (different color lines), keeping the initial positions and velocities of the stars are fixed, see Fig.~\ref{fig:imf-col}. For the second 10 random sets we used one selected initial muss function and randomize the stars positions and velocities (different color lines), see Fig.~\ref{fig:coo-col}. On the above figures we present as a black thick line the observed cumulative number distribution of stars for both clusters. The total cumulative number distribution of stars including the stellar background are presented on the figures as a constantly growing gray line. Because the observations have own limitations due to the Gaia satellite specifications, we select from the numerical models only the stars which are inside the specific stellar mass range - from 0.28 to 4.0 M$_\odot$. In the our comparison figures we also exclude the neutron stars and black holes. The dotted gray lines on the figures represents the one $\pm \sigma$ difference levels from observed cumulative number distribution of stars.

As we can see from figures Fig.~\ref{fig:imf-col} and Fig.~\ref{fig:coo-col}
both sets of randomization of our best fitted physical model for clusters Cr 135 and UBC 7 are well inside the one $\pm \sigma$ gap.  

\subsection{Bibliography}

As with standard \LaTeX, there are two ways of producing a bibliography;
either by compiling a list of references by hand (using a
\verb"thebibliography" environment), or by using BibTeX with a suitable
bibliographic database with the bibliography style provided with the iauguide.tex like \verb"\bibliographystyle{iaulike}". The iau.bst will produce the bibliography which is similar to IAU style but not exactly. If any modification has to be made with iau.bst can be adjusted during manuscript preparation but the updated bst file should be given with source files. However, contributors are encouraged to format
their list of references style outlined in section~\ref{fullref}
below.

\subsubsection{References in the text}

References in the text are given by author and date.
Whichever method is used to produce the bibliography, the references in
the text are done in the same way. Each bibliographical entry has a key,
which is assigned by the author and used to refer to that entry in the
text. There is one form of citation -- \verb"\cite{key}" -- to produce the
author and date. Thus, \cite{2013EAS....62..143B} is produced~by
\begin{verbatim}
  \cite{2013EAS....62..143B}.
\end{verbatim}

\verb"natbib.sty" is common package to handle various reference and its cross citations. The natbib descriptions are available in the document can be find in the web link \verb"https://ctan.org/pkg/natbib?lang=en"

To avoid this, the function of the \verb"\cite" command's optional argument
has been changed. For example, the \verb"\cite" command for the
`\verb"mcc90"' entry gives:
\[ \hbox{McCord (1990)} \]
but you want the following to appear in the text:
\[ \hbox{McCord (1990, see p.~119)} \]
you would then use:
\[ \hbox{\verb"\cite[McCord (1990), see p.~119)]{mcc90}"} \]
to obtain the desired result. Notice that you have to supply
the round brackets as well in the optional argument.

\bibliographystyle{iaulike}  
\bibliography{iau_ishchenko}



\end{document}